\documentstyle[sprocl,epsf]{article}

\newcommand{\bra}[1]{\left< #1 \right|}                 
\newcommand{\ket}[1]{\left| #1 \right>}                 

\begin{document}

\title{Full QCD with Wilson Fermions: Recent Results from the SESAM Collaboration}

\author{{\sf SESAM-Collaboration}\footnote{Talk given by H.~Hoeber}\\
N.~Eicker, U.~Gl\"assner, S.~G\"usken, H.~Hoeber, Th.~Lippert,
G.~Ritzenh\"ofer, K.~Schilling, G.~Siegert, A.~Spitz, P.~Ueberholz and J.~Viehoff.}

\address{BUGH Wuppertal and HLRZ-J\"ulich-DESY-Hamburg}

\maketitle\abstracts{We present recent results of SESAM's large scale
lattice simulation of QCD with two dynamical flavours of Wilson fermions. 
}
\section{Introduction}
As was pointed out by Gottlieb \cite{Gott}, simulations with 
dynamical Wilson fermions at weak coupling on large lattices and at light 
quark masses are still lacking. In an effort to improve on this
situation, SESAM (Sea-Quark Effects on Spectrum and Matrix Elements)
and T${\chi}$L (Towards the Chiral Limit \cite{TxL}) have been producing dynamical 
gauge configurations (with Wilson fermions) at a total of 4 sea-quark
mass values, on two lattice volumes and working at a coupling which -
as we shall show below - corresponds to a quenched coupling in the
scaling regime. In this short note, we summarize some of SESAM's
recent findings; more details can be found in
\cite{tech,SESpot,SES96,Glaessner,SESflav}. 
\section{Status of Simulation}
We work with two dynamical Wilson fermions on a lattice of dimensions
$16^3 \times 32$ and at a strong coupling of $\beta = 5.6$. SESAM is
now close to completing its QH2 ($4\times 8 \times 8$ nodes) 
run-time of approx. 350 days and in this
time we have produced, after thermalisation, 5000 trajectories of unit
length ($100 \pm 20$ molecular dynamics steps with $dt = 0.01$ - see
\cite{tech} for details) at sea quark values $\kappa_{\rm sea} =
\{0.156, 0.157, 0.1575\}$. A
half-time analysis was presented at Lattice 96 \cite{SES96} where we
analysed $\{100,160,100\}$ configurations per sea-quark taken at intervals of 25
units. This interval size is motivated from clean signals in the
autocorrelation function which emerge once the ensembles become
larger than  aprrox. 2000 trajectories. 
As an example we quote the integrated autocorrelation
times of the plaquette $\tau_{int}^{\rm pla} =
\{3.3(5),4.1(3),7.1(5)\}$ and - as a ``worst case'' - that of the average number of
iterations in the HMC (using BiCGStab) $\tau_{int}^{\rm N_{\rm it}} = \{20(1),27(3),31(4)\}$
\cite{Gero}.
\section{Spectrum}
For each sea-quark mass we calculate hadron masses with valence quarks
$\kappa_{\rm val} = \{0.1555, 0.1560, 0.1565, 0.1570, 0.1575
\}$. Gauge invariant smearing (50 iterations and smearing parameter $\alpha$=4)
is applied to obtain smeared sources and smeared/local sinks. We
use uncorrelated single-exp. fits to SL and SS data (simultaneously).
Least-$\chi^2$ fits favour (linear $+$ quadratic)
chiral parametrizations \cite{Gasser} for pseudoscalar and vector mesons as well as the nucleon
(for vector particle and nucleon, a term $\propto m^{3/2}$ does equally well
\cite{Glaessner}). 
\begin{table}[hbt]
\begin{center}
\begin{tabular}{|c|ccccc|}
\hline
$\kappa_{\rm sea}$ & 
$\kappa_c $ & $am_{\rho}$ & $a^{-1}_{\rho}[GeV]$ & $\kappa_{\rm
  light}$ & $\kappa_{\rm strange}$ \\ \hline 
0.156  &  0.16065(8) & 0.359(8) & 2.14(5)  & 0.16058(8) & 0.1576(2) \\ 
0.157  &  0.15987(6) & 0.341(8)  & 2.23(7) & 0.15980(5) & 0.1569(2) \\ 
0.1575 &  0.15963(11) & 0.316(10) &2.44(8) & 0.15944(7) & 0.1544(1) \\ \hline 
quenched &   & & & & \\ \hline
$\beta =6.0$ &  0.15718(6) &0.330(8)  & 2.33(6)  & 0.15709(4) & 0.1544(1) \\ \hline
\end{tabular}
\caption{\label{fitparams}Lattice spacings and $\kappa$-values.
}
\end{center}
\end{table}
The lattice spacing is determined using the rho mass at $\kappa_c$;
  using instead $\kappa_{\rm
  light}$ reduces $a^{-1}$ by no more than 1\%. The values for
  $\kappa_{\rm light}$ and $\kappa_{\rm strange}$ are obtained by 
interpolating to the mass ratios ${m^2_{\pi} \over m^2_{\rho}}$ and
  ${m_{\phi} \over m_{\rho}}$ and can be used to extract the quark
  masses \cite{gupta96}. From table \ref{fitparams} we note that
  our lattice spacings correspond to a quenched $\beta$ of around 6.0,
  one that is at the onset of the scaling regime. These values of
  $a^{-1}$ are in
  agreement with those extracted from the interquark-force\cite{SESpot}.
\par
An alternative way of taking the chiral limit is to use only data with
$\kappa_{\rm sea} = \kappa_{\rm val}$ in the extrapolation of
mass ratios. We postpone this discussion until 
higher statistics are reached and an additional $\kappa_{\rm sea}$
from T$\chi$L becomes available.
\section{The $\pi$-Nucleon $\sigma$ term}
Over the past years there has been significant {\it
computational} progress in the calculation of flavour singlet matrix
elements (see \cite{Okawa} and references therein). It was shown, in
particular, that the noisy-estimator technique with a random $Z_2$ noise
is a promising method to calculate the amplitude of the disconnected
contribution to  $\sigma_{\pi N}$, where the term ``disconnected''
refers to :
\begin{eqnarray}
\sigma_{\pi N} &=& m_q \bra N \bar u u + \bar d d - 2 \bar s s \ket N + 
2 m_q \bra N \bar s s \ket N \nonumber \\
&=& \sigma_{\pi N}^{\rm connected} +  \sigma^{\rm disconnected}_{\pi
N}\, .
\end{eqnarray}
Here, $\ket N$ is a nucleon state and $m_q$ denotes the current quark
mass of the $u$ and $d$ (taken to be equal). 
\par
\begin{figure}[tb]
\begin{center}
\noindent\parbox{12.0cm}{
\parbox{4.cm}{\epsfxsize=4.cm\epsfbox{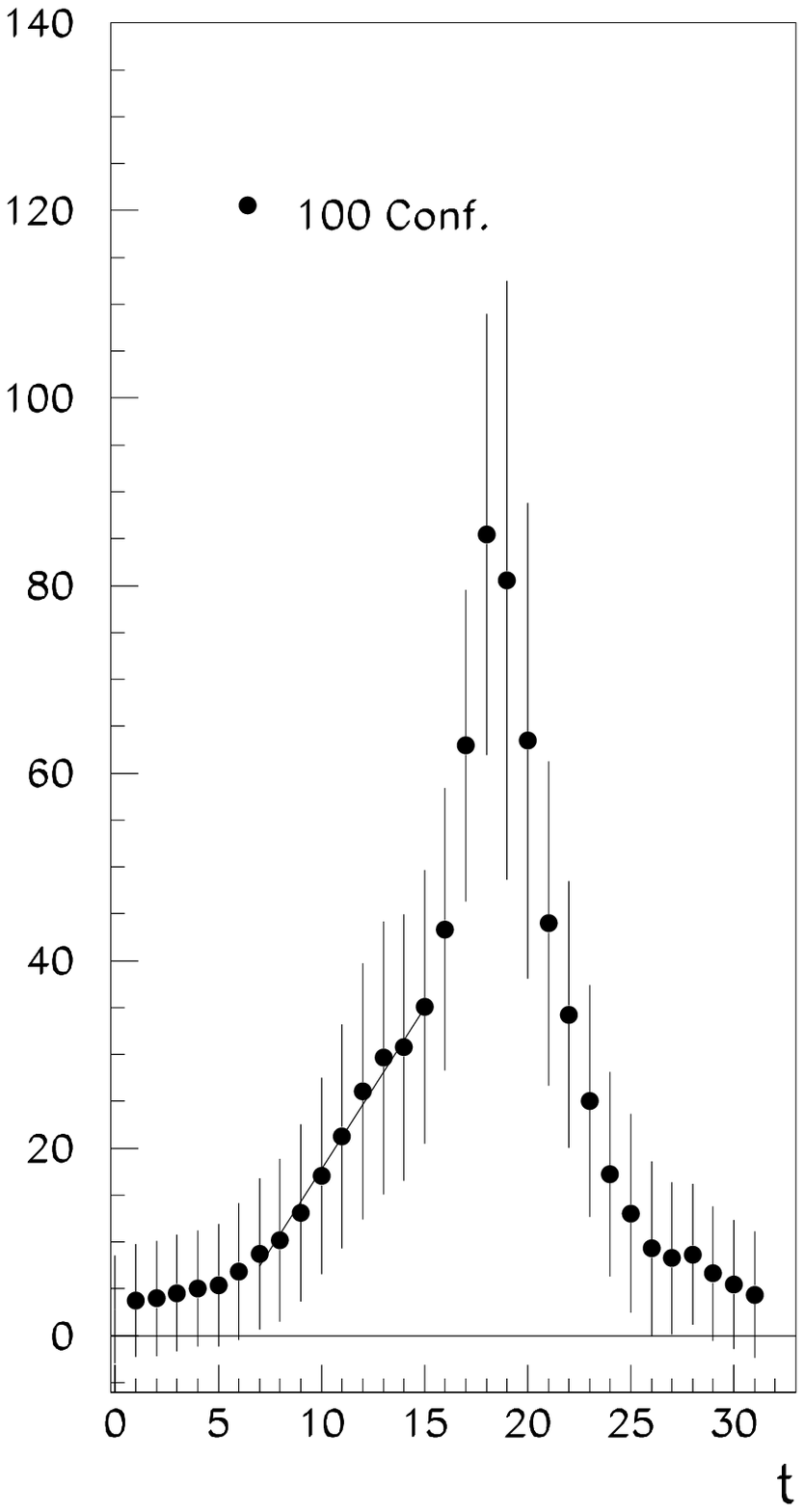}}\nolinebreak
\parbox{4.cm}{\epsfxsize=4.cm\epsfbox{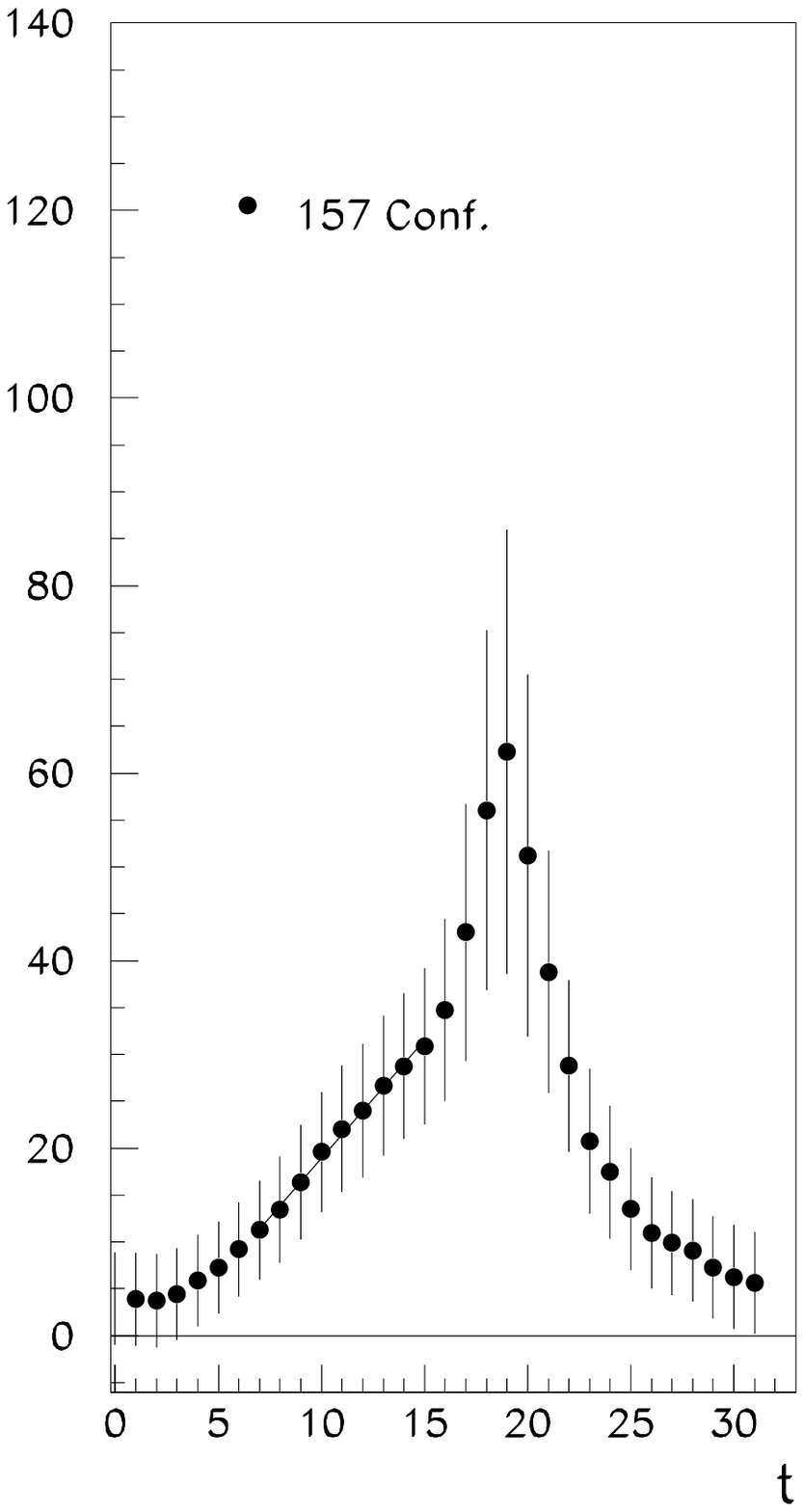}}\nolinebreak
\parbox{4.cm}{\epsfxsize=4.cm\epsfbox{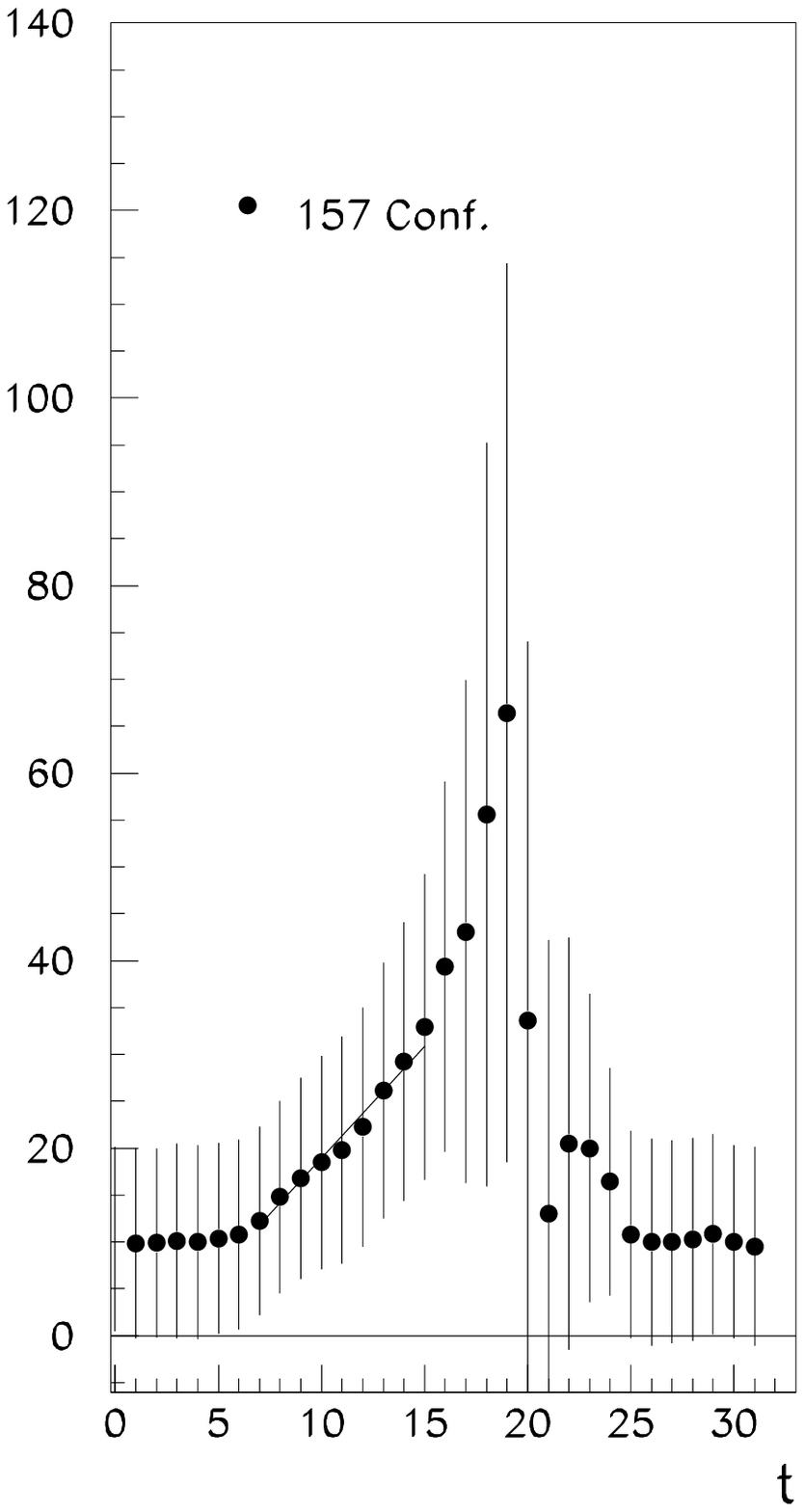}}}
\end{center}
\caption[a]{\label{plot1}
\it $R(t)$ measured with the noisy estimator technique($N_E=300$) on 100 and 157
gau\-ge field configurations. Linear fits were performed
in the range $t = 7$ to $t =  15$. The furthermost plot to the right
shows a simulation with 157 {\sf quenched} gauge-configurations
($\beta = 6.0$ at $\kappa \approx \kappa_{\rm strange}$).}
\end{figure}
\begin{figure}[tb]
\begin{center}
\noindent\parbox{12.0cm}{
\parbox{5.7cm}{\epsfxsize=5.7cm\epsfbox{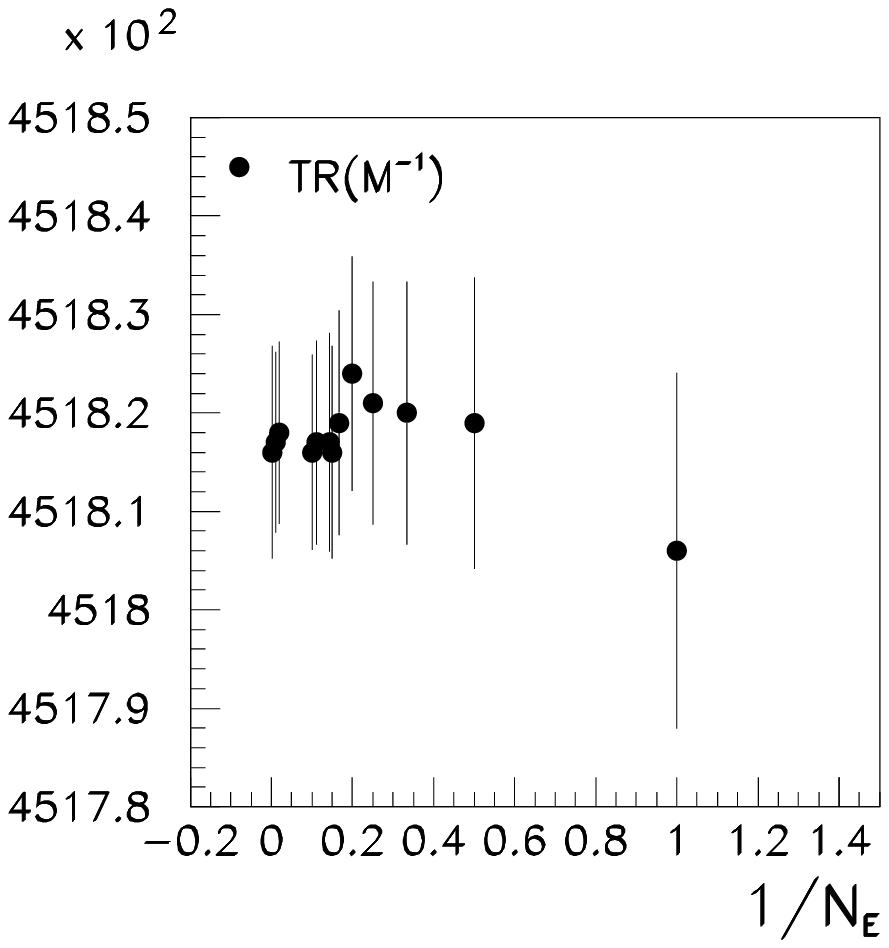}}\nolinebreak
\parbox{5.7cm}{\epsfxsize=5.7cm\epsfbox{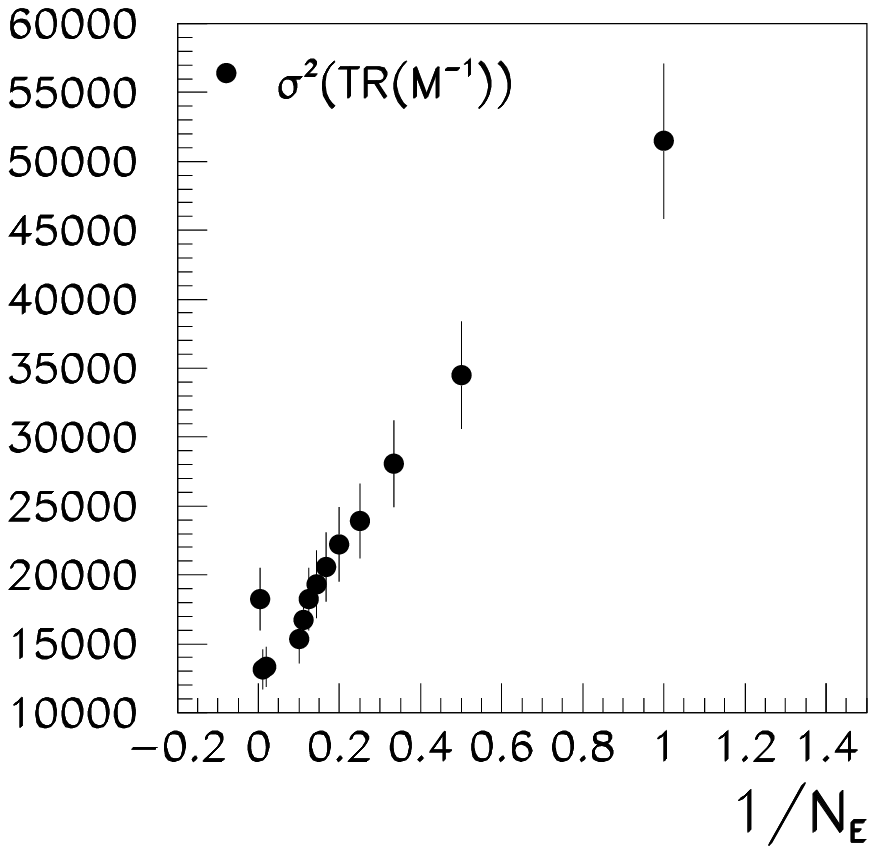}} \\
}
\end{center}
\caption[a]{\label{plot2}
\it $Tr M^{-1}$ and its variance as a
function of $1/N_E$ (157 gauge configurations).}  
\end{figure}

Pioneering attempts to calculate the disconnected amplitude in full QCD have not
led to overly clear signals \cite{altmeyer}. Quenched (high-statistics) calculations
have produced rather more promising data (see \cite{Okawa} for a review
  and \cite{Dong,Fukugita,liu}), however, it is unclear, a
priori, what meaning can be attributed to disconnected amplitudes for
gauge configurations which disregard the effects of dynamical
quarks. In addition, quark masses are found to be much smaller in full
lattice QCD adding a significant uncertainty to the quenched
calculations\cite{gupta96}. \\
\underline{{\bf Algorithmic Investigation}}\\
We have carried out an extensive algorithmic investigation for the
calculation of the disconnected diagram using our full QCD
configurations to check whether - and if so, at what cost - 
a decent signal can be obtained 
on a sample of maximally 200 configurations \cite{SESflav}. The matrix
element of the nucleon with $\bar q q$ insertions is obtained on
the lattice from the following ratio :
\begin{eqnarray}
{\cal R}(t) = { \sum_{\vec x} \bra{N(\vec x,t)} \sum_z \bar q q (z) \ket{N(0)}
\over \sum_{\vec x}\left<{N(x)}\right. | \left. {N(0)} \right> } &=& -{\partial \over \partial m}
{\sf ln}\, \Delta^{-1}(t) \nonumber \\ 
&\stackrel{t \gg 1}{\rightarrow}& {\sf constant} + t \, \bra N \bar q
q \ket N \,,
\end{eqnarray}
where $\Delta(t)$ is the nucleon propagator. 
The hadron correlator is calculated in a standard fashion (using
smearing as above). The disconnected contribution to the three-point function in the numerator is
given by the correlation of the nucleon propagator and the disconnected
insertion  $\sum_x \left({\sf Tr}\, \Delta_{xx} \right)$. 
In the following we present the results of our numerical
investigation for the most efficient calculation of such disconnected
diagrams.\\
\underline{{Noisy Estimator Techniques :}} The noisy estimator
technique uses complex random sources with the 
property $\lim_{N_E \rightarrow \infty}
{1 \over N_E} \sum^{N_E}_{1} \eta_i^*(E,C)\eta_j(E,C) = \delta_{i,j}
$ to calculate $(\eta^{\dagger}M^{-1}\eta)$ $N_E$ times per configuration
$C$. Using 157 configurations at our intermediate sea quark mass
(i.e. ${m_{\pi} \over m_{\rho}} = 0.76(1)$ and $m_q \simeq 1.3\, m_s$)
we have first compared {\it Gaussian} and {\it $Z_2$} noise sources to
calculate the chiral condensate $\chi \propto {\sf
Tr}(M^{-1})$. Monitoring the standard deviation of $\chi$ versus the
number of estimates $N_E$ we found $Z_2$ to outdo Gaussian noise by a
near factor of 2; we pushed $N_E$ all the way up to 300 in this
investigation. Next, we varied the accuracy of the inversion residual
$r = {||M x  - \phi || \over ||x||}$ which, obviously, we wish to relax
as much as possible. Monitoring the quantity $\delta \chi = \chi(r=
10^{-5}) - \chi(r)$ we find that we can choose $r \simeq 10^{-4}$ and
be safely within the $1\, \sigma$ error margin of the $Z_2$ technique
($N_E = 300$). We now turn to the quality of the signal for
${\cal R}(t)$ which can be obtained with our 157 dynamical configurations and to
the dependence of the signal-quality on the number of estimates applied;
recall that so far, we have chosen $N_E = 300$, a number suggested by
the authors of ref.\cite{liu}. In figure \ref{plot1} we show ${\cal
R}(t)$ measured on 100 and 157 configurations. With more than 100
configurations the signal becomes acceptable and the error starts to
display a ${1 \over \sqrt{N_c}}$ behaviour. For comparison we show a
plot from a quenched simulation with the same number of configurations
(157); note that the signal is much worse, indicating that the
determinant in the probability density with which configurations are
sampled has a smoothing effect.
Since a decent
signal emerges, we monitor in figure \ref{plot2} 
the chiral condensate and its variance as a function of ${1
\over N_E}$. 
Whereas the mean value of ${\cal R}(t)$ is unaffected
when varying $N_E$, figure \ref{plot2} shows clearly that it does not pay to
increase $N_E$ beyond 10-20 ! This drastically reduces the cost of the
calculation. An analogous study using the volume source technique \cite{Fukugita},
where $M(C) x = \phi$ is solved for a volume source vector $\phi_i =
1$, shows a much worse signal. Our method of choice is therefore the
stochastic estimator technique with about 20 $Z_2$ sources per configuration
and a relaxed residual $r = 10^{-4}$. The results of our simulation,
where we calculate connected and disconnected amplitudes in full and
quenched QCD will be presented in a forthcoming paper.

\end{document}